\documentclass[citeautoscript,floatfix,aps,prx,onecolumn,superscriptaddress]{revtex4-2}

\usepackage{gensymb}
\usepackage{graphicx}
\usepackage{subcaption}
\usepackage{bm}
\usepackage{float}
\usepackage{amsmath}
\usepackage[
  unicode,
  pdfencoding=auto,
  psdextra,
  bookmarks=true,
  bookmarksopen=true,
  bookmarksnumbered=true,
  colorlinks=true,
  linkcolor=blue,
  citecolor=blue,
  urlcolor=blue,
  breaklinks=true
]{hyperref}

\usepackage{listings}
\usepackage{xcolor}

\newcommand{\pf}{\mathrm{pf}}

\begin{document}

\title{lrux: Fast low-rank updates of determinants and Pfaffians in JAX}

\author{Ao Chen}
\email{chenao.phys@gmail.com}
\affiliation{Division of Chemistry and Chemical Engineering, California Institute of Technology, Pasadena, California 91125, USA}
\affiliation{Center for Computational Quantum Physics, Flatiron Institute, New York 10010, USA}

\author{Christopher Roth}
\affiliation{Center for Computational Quantum Physics, Flatiron Institute, New York 10010, USA}

\begin{abstract}
We present \texttt{lrux}, a JAX-based software package for fast low-rank updates of determinants and Pfaffians, targeting the dominant computational bottleneck in various quantum Monte Carlo (QMC) algorithms. The package implements efficient low-rank updates that reduce the cost of successive wavefunction evaluations from $\mathcal{O}(n^3)$ to $\mathcal{O}(n^2k)$ when the update rank $k$ is smaller than the dimension $n$ of matrices. Both determinant and Pfaffian updates are supported, together with delayed-update strategies that trade floating-point operations for reduced memory traffic on modern accelerators. \texttt{lrux} natively integrates with JAX transformations such as JIT compilation, vectorization, and automatic differentiation, and supports both real and complex data types. Benchmarks on GPUs demonstrate up to $1000\times$ speedup at large matrix sizes. \texttt{lrux} enables scalable, high-performance evaluation of antisymmetric wavefunctions and is designed as a drop-in component for a wide range of QMC workflows. \texttt{lrux} is available at \href{https://github.com/ChenAo-Phys/lrux}{https://github.com/ChenAo-Phys/lrux}.
\end{abstract}

\maketitle

\section{Introduction} \label{sec:intro}

Quantum Monte Carlo (QMC) methods are among the most powerful approaches for simulating interacting electronic systems that cannot be solved exactly. They play a central role in accurately modeling the chemical properties of large molecules and in understanding the collective behavior of electrons in correlated materials. A wide range of QMC algorithms exists, including projective methods such as diffusion Monte Carlo (DMC) \cite{Anderson-DQMC,Ceperly-DQMC}, auxiliary-field QMC (AFQMC) \cite{Hirsch-AFQMC,Blankenbecler-AFQMC,Zhang-CP-AFQMC}, and variational approaches \cite{McMillan-VMC,Foulkes-VMC}, including the recently introduced neural quantum states (NQS) \cite{Carleo-NQS}. Despite their algorithmic diversity, all QMC methods rely on repeatedly sampling from a many-electron wavefunction,         
\begin{equation}
   \psi({\bf x}_1, {\bf x}_2 .... {\bf x}_n ),
\end{equation}
or closely related quantities such as ratios or overlaps between wavefunctions.

A defining feature of electrons is their fermionic antisymmetry under particle exchange, $\psi({\bf x}_1 , {\bf x}_2) = -\psi({\bf x}_2, {\bf x}_1)$, implying that many-electron wavefunctions transform under the alternating representation of the symmetric group. For non-interacting electrons, the minimal representation is a Slater determinant, which can be evaluated in $\mathcal{O}({n^3})$ time for $n$ electrons. A more general antisymmetric form known as the Pfaffian, which naturally represents paired phases \cite{Lou-Superfluid,Chen-HFPS,Roth-Superconductivity}, also scales as $\mathcal{O}({n^3})$ but typically with a larger prefactor. Efficient evaluation of determinants and Pfaffians is therefore a critical bottleneck for scaling QMC calculations to large systems.

In many QMC algorithms, successive configurations differ only by the occupation or position of a small number of orbitals. In this case, determinants and Pfaffians can be updated using low-rank update (LRU) techniques, reducing the computational cost to $\mathcal{O}(n^2 k)$ where $k$ is the rank of the update. Therefore, such updates can provide massive speedups for computing wavefunction ratios and overlaps during Monte Carlo sampling. Moreover, LRU can be used in fermionic NQS \cite{Chen-thesis, Robledo-HFDS, Clark-Backflow, Chen-HFPS, Pfau-FermiNet, Hermann-QuantumChem, Roth-Superconductivity} (as done in \cite{Chen-HFPS, Roth-Superconductivity, Chen-thesis}), provided that the neural-network backflow transformation of the orbitals itself admits a low-rank representation. In this case, the effective update rank is given by the sum of the ranks associated with orbital changes and backflow transformations. 

In this work, we introduce \texttt{lrux}, a JAX-based software package that implements LRUs for determinants and Pfaffians, designed for seamless integration into a wide range of QMC algorithms. JAX \cite{JAX} enables automatic parallelization and just-in-time (JIT) compilation, allowing \texttt{lrux} to efficiently utilize modern GPU architectures, where dense linear-algebra operations can achieve order-of-magnitude speedups over CPU implementations. In addition, \texttt{lrux} supports delayed updates, which trade increased floating-point operations for reduced memory movement, allowing users to optimize performance on their hardware. 

\texttt{lrux} is available at \href{https://github.com/ChenAo-Phys/lrux}{https://github.com/ChenAo-Phys/lrux} and may be installed by \texttt{pip install lrux}. At the time of writing, it is at version 0.1.2 and compatible with \texttt{jax>=0.4.4}.

We provide illustrative examples demonstrating how to use \texttt{lrux} in practical QMC workflows, along with recommended environment settings and dependencies. In particular, enabling double-precision arithmetic is strongly advised to mitigate numerical error accumulation and maintain stability in large-scale simulations. The code below should be put at the top of all example codes we present in this paper.

\begin{lstlisting}[language=Python]
import jax

# Enable double precision in JAX
jax.config.update("jax_enable_x64", True)

import jax.numpy as jnp
import jax.random as jr
import lrux

KEY = jr.key(42)

def get_key():
    """
    A function for generating jax keys. 
    Not intended for general usage.
    """
    global KEY
    KEY, new_key = jr.split(KEY, 2)
    return new_key
\end{lstlisting}

\section{Determinant}

\subsection{Low-rank update of determinant}

Consider consecutive low-rank updates of $n \times n$ matrices $\mathbf{A}_0 \rightarrow \mathbf{A}_1 \rightarrow ... \rightarrow \mathbf{A}_t \rightarrow ...$. At each step, the low-rank update can be expressed as $\mathbf{A}_t = \mathbf{A}_{t-1} + \mathbf{v}_t \mathbf{u}_t^T$, where $\mathbf{u}_t$ and $\mathbf{v}_t$ are $n \times k$ matrices with $k \ll n$. Utilizing the matrix determinant lemma, we have
\begin{equation}
    \det(\mathbf{A}_t) = \det (\mathbf{A}_{t-1} + \mathbf{v}_t \mathbf{u}_t^T)
    = \det (\mathbf{A}_{t-1}) \det (\mathbf{1} + \mathbf{u}_t^T \mathbf{A}_{t-1}^{-1} \mathbf{v}_t),
\end{equation}
and the ratio between two determinants is
\begin{equation} \label{eq:det_ratio}
    r_t = \frac{\det \mathbf{A}_t}{\det \mathbf{A}_{t-1}} = \det \mathbf{R}_t,
\end{equation}
where
\begin{equation} \label{eq:det_R}
    \mathbf{R}_t = \mathbf{1} + \mathbf{u}_t^T \mathbf{A}_{t-1}^{-1} \mathbf{v}_t
\end{equation}
is a $k \times k$ matrix. If $\det \mathbf{A}_{t-1}$ and $\mathbf{A}_{t-1}^{-1}$ have been computed and stored in memory, one can compute $\det \mathbf{A}_1$ with a reduced $\mathcal{O}(n^2k)$ complexity instead of the original determinant complexity $\mathcal{O}(n^3)$. Here, we provide a sample code to compute a rank-1 (k=1) update.

\begin{lstlisting}[language=Python]
n = 4
k = 1

A = jr.normal(get_key(), (n, n))
Ainv = jnp.linalg.inv(A)
detA = jnp.linalg.det(A)

# Random update
u = jr.normal(get_key(), (n, k))
v = jr.normal(get_key(), (n, k))

r = lrux.det_lru(Ainv, u, v)
detA *= r
det_exact = jnp.linalg.det(A + v @ u.T)
assert jnp.isclose(detA, det_exact)
\end{lstlisting}

In many applications, such as continuously computing the ratio of wavefunctions in a Monte-Carlo chain, it is necessary to perform several successive low-rank updates. This requires keeping track of the matrix inverse, $\mathbf{A}_t^{-1}$, which can be updated via the Sherman–Morrison formula,
\begin{equation} \label{eq:det_inv}
    \mathbf{A}_t^{-1} = (\mathbf{A}_{t-1} + \mathbf{v}_t \mathbf{u}_t^T)^{-1}
    = \mathbf{A}_{t-1}^{-1} 
    - \mathbf{A}_{t-1}^{-1} \mathbf{v}_t \mathbf{R}_t^{-1} \mathbf{u}_t^T \mathbf{A}_{t-1}^{-1},
\end{equation}
which requires $\mathcal{O}(n^2 k)$ complexity instead of the $\mathcal{O}(n^3)$ needed to compute the inverse from scratch. By repeating the procedure of computing the ratio between wavefunctions and updating the stored inverse, one can obtain consecutive low-rank updates. The memory complexity of this algorithm is $\mathcal{O}(n^2)$ for storing $\mathbf{A}_t^{-1}$. We show below the sample code for consecutive updates, still for the rank-1 row update case.

\begin{lstlisting}[language=Python]
n = 4
k = 1

A = jr.normal(get_key(), (n, n))
Ainv = jnp.linalg.inv(A)
detA = jnp.linalg.det(A)

det_lru_fn = jax.jit(lrux.det_lru, static_argnums=3, donate_argnums=0)

# Random update
u1 = jr.normal(get_key(), (n, k))
v1 = jr.normal(get_key(), (n, k))
r, Ainv = det_lru_fn(Ainv, u1, v1, return_update=True)
detA *= r

# Random update
u2 = jr.normal(get_key(), (n, k))
v2 = jr.normal(get_key(), (n, k))
r, Ainv = det_lru_fn(Ainv, u2, v2, return_update=True)
detA *= r

det_exact = jnp.linalg.det(A + v1 @ u1.T + v2 @ u2.T)
assert jnp.isclose(detA, det_exact)
\end{lstlisting}

The numerical stability of Eq.\,\eqref{eq:det_ratio} relies on the non-singularity of $\mathbf{A}_{t-1}^{-1}$, indicating $\det\mathbf{A}_{t-1} \neq 0$, otherwise the ratio $r_t$ cannot be computed. Similarly, Eq.\,\eqref{eq:det_inv} requires non-singular $\mathbf{R}_t$ and $\det \mathbf{R}_t \neq 0$, otherwise $\mathbf{A}_t$ is singular and non-invertible. Therefore, we suggest that users enable double precision to avoid occasional numerical (near-)singularity in practice.

\subsection{Matrix updates}

In the following, we show that many matrix updates, including the common operation of single-row and single-column updates in QMC, can be written into a unified low-rank expression $\mathbf{A}_1 - \mathbf{A}_0 = \mathbf{vu}^T$. Here, we list several special cases of low-rank updates and the corresponding codes for expressing $\mathbf{u}$ and $\mathbf{v}$ in \texttt{lrux}.

\begin{itemize}
    \item \textbf{Rank-1 row update:}
        \begin{equation} \label{eq:det_rank1_row}
            \mathbf{A}_1 - \mathbf{A}_0 = \begin{pmatrix}
                0 & 0 & 0 & 0 \\ 
                u_0 & u_1 & u_2 & u_3 \\
                0 & 0 & 0 & 0 \\ 
                0 & 0 & 0 & 0 \\ 
            \end{pmatrix}
            = \begin{pmatrix}
                0 \\  1 \\ 0 \\ 0
            \end{pmatrix}
            (u_0, u_1, u_2, u_3)
            = \mathbf{vu}^T.
        \end{equation}
        
\begin{lstlisting}[language=Python]
u = jnp.array([u0, u1, u2, u3])
v = 1  # one-hot vector non-zero at position 1
\end{lstlisting}

    \item \textbf{Rank-1 column update:}
        \begin{equation}
            \mathbf{A}_1 - \mathbf{A}_0 = \begin{pmatrix}
                0 & 0 & v_0 & 0 \\ 
                0 & 0 & v_1 & 0 \\
                0 & 0 & v_2 & 0 \\ 
                0 & 0 & v_3 & 0 \\ 
            \end{pmatrix}
            = \begin{pmatrix}
                v_0 \\ v_1 \\ v_2 \\ v_3
            \end{pmatrix}
            (0, 0, 1, 0)
            = \mathbf{vu}^T.
        \end{equation}
        
\begin{lstlisting}[language=Python]
u = 2  # one-hot vector non-zero at position 2
v = jnp.array([v0, v1, v2, v3]) 
\end{lstlisting}
        
    \item \textbf{Rank-2 row update:}
        \begin{equation}
            \mathbf{A}_1 - \mathbf{A}_0 = \begin{pmatrix}
                0 & 0 & 0 & 0 \\ 
                u_{00} & u_{01} & u_{02} & u_{03} \\
                0 & 0 & 0 & 0 \\ 
                u_{10} & u_{11} & u_{12} & u_{13} \\
            \end{pmatrix}
            = \begin{pmatrix}
                0 & 0 \\ 1 & 0 \\ 0 & 0 \\ 0 & 1
            \end{pmatrix}
            \begin{pmatrix}
                u_{00} & u_{01} & u_{02} & u_{03} \\
                u_{10} & u_{11} & u_{12} & u_{13} \\
            \end{pmatrix}
            = \mathbf{vu}^T.
        \end{equation}
        
\begin{lstlisting}[language=Python]
u = jnp.array([[u00, u01, u02, u03], [u10, u11, u12, u13]]).T

# two concatenated one-hot vectors non-zero at position 1 and 3
v = jnp.array([1, 3])
\end{lstlisting}

    \item \textbf{Simultaneous update of row and column:}
        \begin{equation}
            \mathbf{A}_1 - \mathbf{A}_0 = \begin{pmatrix}
                0 & 0 & v_0 & 0 \\ 
                u_0 & u_1 & u_2 + v_1 & u_3 \\
                0 & 0 & v_2 & 0 \\ 
                0 & 0 & v_3 & 0 \\ 
            \end{pmatrix}
            = \begin{pmatrix}
                0 & v_0 \\ 1 & v_1 \\ 0 & v_2 \\ 0 & v_3
            \end{pmatrix}
            \begin{pmatrix}
                u_0 & u_1 & u_2 & u_3 \\
                0 & 0 & 1 & 0 \\
            \end{pmatrix}
            = \mathbf{vu}^T.
        \end{equation}
        
\begin{lstlisting}[language=Python]
xu = jnp.array([u0, u1, u2, u3])
eu = 2
# concatenation of dense and one-hot representations
u = (xu, eu)

xv = jnp.array([v0, v1, v2, v3])
ev = 1
# concatenation of dense and one-hot representations
v = (xv, ev)
\end{lstlisting}
        
\end{itemize}

We suggest that users express $\mathbf{u}$ and $\mathbf{v}$ with one-hot representations whenever possible, as it sometimes greatly accelerates the computation. For instance, when one computes $\mathbf{u}^T \mathbf{Mv}$ with $\mathbf{u}$ and $\mathbf{v}$ given in Eq.\,\eqref{eq:det_rank1_row}, the naive dense operation \texttt{u @ M @ v\_dense} has $\mathcal{O}(n^2)$ complexity, while the one-hot operation \texttt{u @ M[:, v\_onehot]} only has $\mathcal{O}(n)$ complexity.

\subsection{Delayed updates}

The time bottleneck of LRU comes from $\mathbf{A}_{t-1}^{-1} \mathbf{v}_t \mathbf{R}_t^{-1} \mathbf{u}_t^T \mathbf{A}_{t-1}^{-1}$ in Eq.\,\eqref{eq:det_inv}, which unavoidably involves the matrix product of $n \times k$ and $k \times n$ matrices. While the floating-point complexity, $\mathcal{O}(n^2k)$, is the same as other parts of the algorithm, the computation is often bounded by memory bandwidth for small $k$. In this situation, a delayed update strategy can be utilized to accelerate the computation. At each time step $t$, we define
\begin{equation} \label{eq:det_a}
    \mathbf{a}_t = \mathbf{A}^{-1}_{t-1} \mathbf{v}_t,
\end{equation}
\begin{equation} \label{eq:det_b}
    \mathbf{b}_t = (\mathbf{A}^{-1}_{t-1})^T \mathbf{u}_t (\mathbf{R}_t^{-1})^T,
\end{equation}
such that
\begin{equation} \label{eq:delayed_Ainv}
    \mathbf{A}_\tau^{-1} = \mathbf{A}_{\tau-1}^{-1} - \mathbf{a}_\tau \mathbf{b}_\tau^T
    = \mathbf{A}_0^{-1} - \sum_{t=1}^\tau \mathbf{a}_t \mathbf{b}_t^T.
\end{equation}
Therefore, instead of explicitly computing $\mathbf{A}_\tau^{-1}$ in each step according to Eq.\,\eqref{eq:det_inv}, one can alternatively store $\mathbf{A}_0^{-1}$, $\mathbf{a}_t$, and $\mathbf{b}_t$, which contains enough information to reconstruct $\mathbf{A}_\tau^{-1}$. Then combining Eq.\,\eqref{eq:det_R} and Eq.\,\eqref{eq:delayed_Ainv}, we can compute $\mathbf{R}_\tau$ as
\begin{equation}
    \mathbf{R}_\tau = \mathbf{1} + \mathbf{u}_\tau^T \mathbf{A}_0^{-1} \mathbf{v}_\tau - \sum_{t=1}^{\tau-1} (\mathbf{u}^T_\tau \mathbf{a}_t) (\mathbf{b}_t^T \mathbf{v}_\tau)
\end{equation}
with $\mathcal{O}(n^2k + \tau n k^2)$ complexity, where the memory-bound computation $\mathbf{a}_t \mathbf{b}_t^T$ is avoided. Similarly, one can combine Eq.\,\eqref{eq:det_a}, Eq.\,\eqref{eq:det_b}, and Eq.\,\eqref{eq:delayed_Ainv} to obtain
\begin{equation}
    \mathbf{a}_\tau = \mathbf{A}_0^{-1} \mathbf{v}_\tau - \sum_{t=1}^{\tau-1} \mathbf{a}_t (\mathbf{b}_t^T \mathbf{v}_\tau),
\end{equation}
\begin{equation}
    \mathbf{b}_\tau = \left[ (\mathbf{A}_0^{-1})^T \mathbf{u}_\tau - \sum_{t=1}^{\tau-1} \mathbf{b}_t ( \mathbf{a}_t^T \mathbf{u}_\tau) \right] (\mathbf{R}_\tau^{-1})^T
\end{equation}
also with $\mathcal{O}(n^2k + \tau n k^2)$ complexity, and the memory-bound computation $\mathbf{a}_t \mathbf{b}_t^T$ is again avoided. The memory complexity for storing $\mathbf{A}_0^{-1}$, $\mathbf{a}_t$, and $\mathbf{b}_t$ is $\mathcal{O}(n^2 + \tau n k)$. To avoid the infinite growth of $\tau$, one can set an upper bound $T$ and reconstruct the full matrix $\mathbf{A}_\tau^{-1}$ using Eq.\,\eqref{eq:delayed_Ainv} when $\tau = T$. Then this $\mathbf{A}_\tau^{-1}$ is set as the new $\mathbf{A}_0^{-1}$, and $\mathbf{a}_t$ and $\mathbf{b}_t$ are set to 0 to start a new round of delayed updates. By controlling $\tau \leq T < n/k$, the overall complexity of delayed updates remains $\mathcal{O}(n^2 k)$. Typically, one chooses $T \approx n/10k$, but the exact optimal value depends strongly on the hardware and matrix size. The sample code of delayed updates is shown below for reference, where we set $T=4$.

\begin{lstlisting}[language=Python]
n = 10
max_delay = 4
A = jr.normal(get_key(), (n, n))
carrier = lrux.init_det_carrier(A, max_delay)
detA = jnp.linalg.det(A)

det_lru_fn = jax.jit(
    lrux.det_lru_delayed, static_argnums=(3, 4), donate_argnums=0
)

for i in range(20):
    current_delay = i % max_delay
    # random update
    u = jr.normal(get_key(), (n,))
    # select a random row
    v = jr.randint(get_key(), shape=(1,), minval=0, maxval=n)  # one-hot representation
    r, carrier = det_lru_fn(carrier, u, v, True, current_delay)
    detA *= r

    # verify the low-rank update result
    A = A.at[v].add(u)
    detA_exact = jnp.linalg.det(A)
    assert jnp.isclose(detA, detA_exact)
\end{lstlisting}

\section{Pfaffian}

The Pfaffian maps a skew-symmetric matrix to a number such that the output is antisymmetric under the exchange of columns or rows, which is often utilized to describe pairings between fermions. One can reference Appendix \ref{sec:Pfaffian} for the definition and important properties of Pfaffians.

\subsection{Low-rank update of Pfaffian}

Consider two $n \times n$ skew-symmetric matrices $\mathbf{A}_t$ and $\mathbf{A}_{t-1}$ that differ only by a low-rank update. Then we have a general expression
\begin{equation}
    \mathbf{A}_t - \mathbf{A}_{t-1} = -\mathbf{u}_t \mathbf{J} \mathbf{u}_t^T,
\end{equation}
where $\mathbf{u}$ is an $n \times 2k$ matrix with $k \ll n$, and
\begin{equation}
    \mathbf{J} = \begin{pmatrix}
        \mathbf{0} & \mathbf{1} \\ -\mathbf{1} & \mathbf{0}
    \end{pmatrix}
\end{equation}
is a $2k \times 2k$ skew-symmetric identity matrix. Utilizing Eq.\,\eqref{eq:pf_low_rank} and $\mathbf{J}^{-1} = -\mathbf{J}$, we have
\begin{equation}
\begin{split}
    \pf \mathbf{A}_t = \pf (\mathbf{A}_{t-1} - \mathbf{u}_t \mathbf{J} \mathbf{u}_t^T) 
    = \frac{\pf \mathbf{A}_{t-1}}{\pf \mathbf{J}}
    \pf ( \mathbf{J} + \mathbf{u}_t^T \mathbf{A}_{t-1} \mathbf{u}_t),
\end{split}
\end{equation}
and the ratio between two Pfaffians is
\begin{equation} \label{eq:pf_ratio}
    r_t = \frac{\pf \mathbf{A}_t}{\pf\mathbf{A}_{t-1}} 
    = \frac{\pf \mathbf{R}_t}{\pf \mathbf{J}},
\end{equation}
where $\pf \mathbf{J} = (-1)^{k(k-1)/2}$ can be derived directly from Eq.\,\eqref{eq:pf_offdiagonal}, and
\begin{equation} \label{eq:pf_R}
    \mathbf{R}_t = \mathbf{J} + \mathbf{u}_t^T \mathbf{A}_{t-1}^{-1} \mathbf{u}_t.
\end{equation}
Therefore, $\pf\mathbf{A}_t$ can be computed with $\mathcal{O}(n^2 k)$ time complexity instead of the original Pfaffian complexity $\mathcal{O}(n^3)$ if $\pf\mathbf{A}_{t-1}$ and $\mathbf{A}_{t-1}^{-1}$ have been computed and stored in memory. The memory complexity is $\mathcal{O}(n^2)$ for storing $\mathbf{A}_{t-1}^{-1}$. We show below a simple sample code for the local update of Pfaffian.
\begin{lstlisting}[language=Python]
n = 4
k = 1

A = jr.normal(get_key(), (n, n))
A = (A - A.T) / 2
Ainv = jnp.linalg.inv(A)
pfA = lrux.pf(A)

# Random update
u = jr.normal(get_key(), (n, 2 * k))

r = lrux.pf_lru(Ainv, u)
pfA *= r
J = lrux.skew_eye(k)
pf_exact = lrux.pf(A - u @ J @ u.T)
assert jnp.isclose(pfA, pf_exact)
\end{lstlisting}

To update $\mathbf{A}_{t-1}^{-1}$ to $\mathbf{A}_t^{-1}$, one needs to utilize the Woodbury matrix identity to obtain
\begin{equation} \label{eq:pf_inv}
    \mathbf{A}_t^{-1} = (\mathbf{A}_{t-1} - \mathbf{u}_t \mathbf{J} \mathbf{u}_t^T)^{-1}
    = \mathbf{A}_{t-1}^{-1} + (\mathbf{A}_{t-1}^{-1} \mathbf{u}_t) \mathbf{R}^{-1} (\mathbf{A}_{t-1}^{-1} \mathbf{u}_t)^T
\end{equation}
where we have used $(\mathbf{A}^{-1}_{t-1})^T = -\mathbf{A}^{-1}_{t-1}$. This update of $\mathbf{A}^{-1}$ has $\mathcal{O}(n^2 k)$ complexity. For faster computation in the $k=1$ case, one can decompose the $n \times 2$ matrix $\mathbf{u}$ into $\mathbf{u} = (\mathbf{x}, \mathbf{y})$ and rewrite Eq.\,\eqref{eq:pf_inv} as
\begin{equation}
\begin{split}
    \mathbf{A}_t^{-1} &= \mathbf{A}_{t-1}^{-1} + (\mathbf{A}_{t-1}^{-1} \mathbf{x}, \mathbf{A}_{t-1}^{-1} \mathbf{y}) 
    \begin{pmatrix}
        0 & r \\ -r & 0
    \end{pmatrix}^{-1} 
    (\mathbf{A}_{t-1}^{-1} \mathbf{x}, \mathbf{A}_{t-1}^{-1} \mathbf{y}) ^T \\
    &= \mathbf{A}_{t-1}^{-1} + \frac{1}{r} (\mathbf{A}_{t-1}^{-1} \mathbf{y}) (\mathbf{A}_{t-1}^{-1} \mathbf{x})^T
    - \frac{1}{r} (\mathbf{A}_{t-1}^{-1} \mathbf{x}) (\mathbf{A}_{t-1}^{-1} \mathbf{y}) ^T. \\
\end{split}
\end{equation}
The \texttt{lrux} code for consecutive LRU with updated matrix inverse $\mathbf{A}^{-1}$ is shown below.
\begin{lstlisting}[language=Python]
n = 4
k = 1

A = jr.normal(get_key(), (n, n))
A = (A - A.T) / 2
Ainv = jnp.linalg.inv(A)
pfA = lrux.pf(A)

det_lru_fn = jax.jit(lrux.pf_lru, static_argnums=2, donate_argnums=0)

# Random update
u1 = jr.normal(get_key(), (n, 2 * k))
r, Ainv = det_lru_fn(Ainv, u1, return_update=True)
pfA *= r

# Random update
u2 = jr.normal(get_key(), (n, 2 * k))
r, Ainv = det_lru_fn(Ainv, u2, return_update=True)
pfA *= r

J = lrux.skew_eye(k)
pf_exact = lrux.pf(A - u1 @ J @ u1.T - u2 @ J @ u2.T)
assert jnp.isclose(pfA, pf_exact)
\end{lstlisting}

\subsection{Matrix updates}

Here, we list several special cases of low-rank updates that are important in applications, and the corresponding codes to express $\mathbf{u}$ in \texttt{lrux}. The one-hot representation of vectors is still recommended whenever possible.

\begin{itemize}
    \item \textbf{Update of 1 row and 1 column}
        \begin{equation} \label{eq:pf_rank2}
            \mathbf{A}_1 - \mathbf{A}_0 = \begin{pmatrix}
                0 & -u_0 & 0 & 0 \\ 
                u_0 & 0 & u_2 & u_3 \\
                0 & -u_2 & 0 & 0 \\ 
                0 & -u_3 & 0 & 0 \\ 
            \end{pmatrix}
            = -\begin{pmatrix}
                u_0 & 0 \\ u_1 & 1 \\ u_2 & 0 \\ u_3 & 0
            \end{pmatrix}
            \begin{pmatrix}
                0 & 1 \\ -1 & 0
            \end{pmatrix}
            \begin{pmatrix}
                u_0 & u_1 & u_2 & u_3 \\
                0 & 1 & 0 & 0\\
            \end{pmatrix}
            = -\mathbf{uJu}^T.
        \end{equation}
        
\begin{lstlisting}[language=Python]
xu = jnp.array([u0, u1, u2, u3])
eu = 1
# concatenation of dense and one-hot representations
u = (xu, eu)
\end{lstlisting}

    \item \textbf{Update of 2 rows and 2 columns}

        \begin{equation}
        \begin{split}
            \mathbf{A}_1 - \mathbf{A}_0 &= \begin{pmatrix}
                0 & -u_{00} & 0 & -u_{10} & 0 & 0\\ 
                u_{00} & 0 & u_{02} & u_{03} - u_{11} & u_{04} & u_{05} \\
                0 & -u_{02} & 0 & -u_{12} & 0 & 0 \\ 
                u_{10} & u_{11} - u_{03} & u_{12} & 0 & u_{14} & u_{15} \\
                0 & -u_{04} & 0 & -u_{14} & 0 & 0 \\ 
                0 & -u_{05} & 0 & -u_{15} & 0 & 0 \\ 
            \end{pmatrix} \\
            &= -\begin{pmatrix}
                u_{00} & u_{10} & 0 & 0 \\
                u_{01} & u_{11} & 1 & 0 \\
                u_{02} & u_{12} & 0 & 0 \\
                u_{03} & u_{13} & 0 & 1 \\
                u_{04} & u_{14} & 0 & 0 \\
                u_{05} & u_{15} & 0 & 0 \\
            \end{pmatrix}
            \begin{pmatrix}
                0 & 0 & 1 & 0 \\
                0 & 0 & 0 & 1 \\
                -1 & 0 & 0 & 0 \\
                0 & -1 & 0 & 0 \\
            \end{pmatrix}
            \begin{pmatrix}
                u_{00} & u_{01} & u_{02} & u_{03} & u_{04} & u_{05} \\
                u_{10} & u_{11} & u_{12} & u_{13} & u_{14} & u_{15} \\
                0 & 1 & 0 & 0 & 0 & 0 \\
                0 & 0 & 0 & 1 & 0 & 0 \\
            \end{pmatrix} \\
            &= -\mathbf{uJu}^T.
        \end{split}
        \end{equation}
        
\begin{lstlisting}[language=Python]
xu = jnp.array([
    [u00, u01, u02, u03, u04, u05], 
    [u10, u11, u12, u13, u14, u15],
]).T
eu = jnp.array([1, 3])
# concatenation of dense and one-hot representations
u = (xu, eu)
\end{lstlisting}

\end{itemize}

\subsection{Delayed updates}

Similar to the LRU of the determinant state, the time bottleneck of the Pfaffian also comes from memory-bound computations. At each step $t$, we define
\begin{equation} \label{eq:pf_a}
    \mathbf{a}_t = \mathbf{A}^{-1}_{t-1} \mathbf{u}_t,
\end{equation}
such that
\begin{equation} \label{eq:pf_delayed_Ainv}
    \mathbf{A}_\tau^{-1} = \mathbf{A}_{\tau-1}^{-1} + \mathbf{a}_\tau \mathbf{R}_\tau^{-1} \mathbf{a}_\tau^T
    = \mathbf{A}_0^{-1} + \sum_{t=1}^\tau \mathbf{a}_t \mathbf{R}_t^{-1} \mathbf{a}_t^T.
\end{equation}
Instead of constructing $\mathbf{A}_\tau^{-1}$ explicitly according to Eq.\,\eqref{eq:pf_inv}, one can alternatively store $\mathbf{A}_0^{-1}$, $\mathbf{a}_t$, and $\mathbf{R}_t^{-1}$ in memory, which contain enough information for reconstructing $\mathbf{A}_\tau^{-1}$. Combining Eq.\,\eqref{eq:pf_R} and Eq.\,\eqref{eq:pf_delayed_Ainv}, we can compute $\mathbf{R}_\tau$ using
\begin{equation}
    \mathbf{R}_\tau = \mathbf{J} + \mathbf{u}_\tau^T \mathbf{A}_0^{-1} \mathbf{u}_\tau
    + \sum_{t=1}^{\tau-1} (\mathbf{u}_\tau^T \mathbf{a}_t) \mathbf{R}_t^{-1} (\mathbf{u}_\tau^T \mathbf{a}_t)^T
\end{equation}
with $\mathcal{O}(n^2 k + \tau n k^2)$ complexity. Then one needs to compute the matrix inverse of the $k \times k$ matrix $\mathbf{R}_\tau$ and store $\mathbf{R}_\tau^{-1}$ in memory for delayed updates. Similarly, by combining Eq.\,\eqref{eq:pf_a} and Eq.\,\eqref{eq:pf_delayed_Ainv}, $\mathbf{a}_\tau$ can be computed as
\begin{equation}
    \mathbf{a}_\tau = \mathbf{A}_0^{-1} \mathbf{u}_t + \sum_{t=1}^{\tau-1} \mathbf{a}_t \mathbf{R}_t^{-1} (\mathbf{u}_\tau^T \mathbf{a}_t)^T
\end{equation}
also with $\mathcal{O}(n^2 k + \tau n k^2)$ time complexity, and the memory complexity is $\mathcal{O}(n^2 + \tau n k)$ for storing $\mathbf{A}_0^{-1}$ and $\mathbf{a}_t$. Similar to the determinant case, one should utilize Eq.\,\eqref{eq:pf_delayed_Ainv} to reconstruct $\mathbf{A}_\tau^{-1}$ and set it as the new $\mathbf{A}_0^{-1}$ when $\tau$ reaches the upper bound $T$. Typically, one chooses $T \approx n/20k$, but the optimal value also depends on the hardware and matrix size. Below, we show the sample code for delayed updates of Pfaffians, where we set $T=4$.

\begin{lstlisting}[language=Python]
n = 10
max_delay = 4
A = jr.normal(get_key(), (n, n))
A = (A - A.T) / 2
carrier = lrux.init_pf_carrier(A, max_delay)
pfA = lrux.pf(A)

det_lru_fn = jax.jit(
    lrux.pf_lru_delayed, static_argnums=(2, 3), donate_argnums=0
)

for i in range(20):
    current_delay = i % max_delay
    # random update
    x = jr.normal(get_key(), (n,))
    # select a random row and column
    e = jr.randint(get_key(), shape=(1,), minval=0, maxval=n)
    u = (x, e)  # combine dense and one-hot representations
    r, carrier = det_lru_fn(carrier, u, True, current_delay)
    pfA *= r

    # verify the low-rank update result
    A = A.at[e[0]].add(x).at[:, e[0]].add(-x)
    pfA_exact = lrux.pf(A)
    assert jnp.isclose(pfA, pfA_exact)
\end{lstlisting}

\section{Benchmark results}

\subsection{One-step comparison}

\begin{figure}[t]
    \centering
    \begin{subfigure}{0.48\textwidth}
        \includegraphics[width=\linewidth]{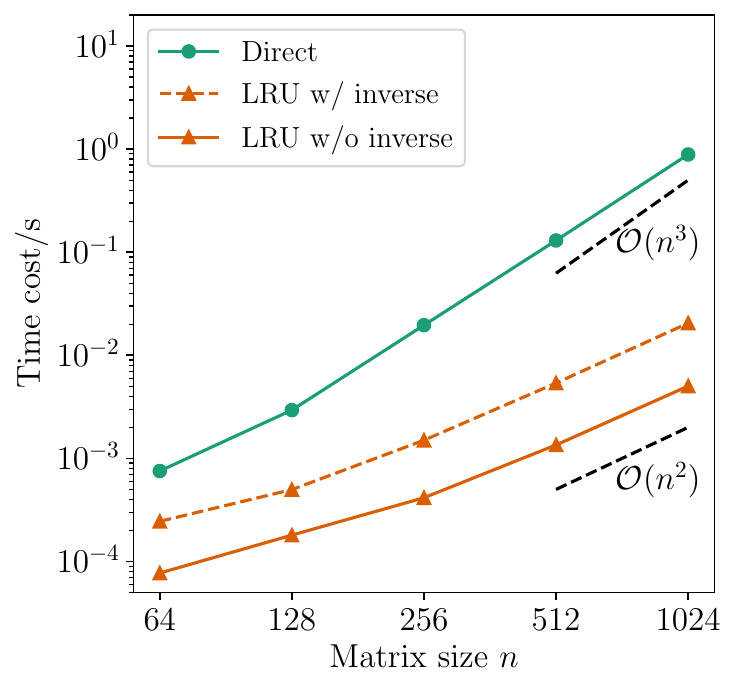}
        \caption{Determinant}
    \end{subfigure}
    \hfill
    \begin{subfigure}{0.48\textwidth}
        \includegraphics[width=\linewidth]{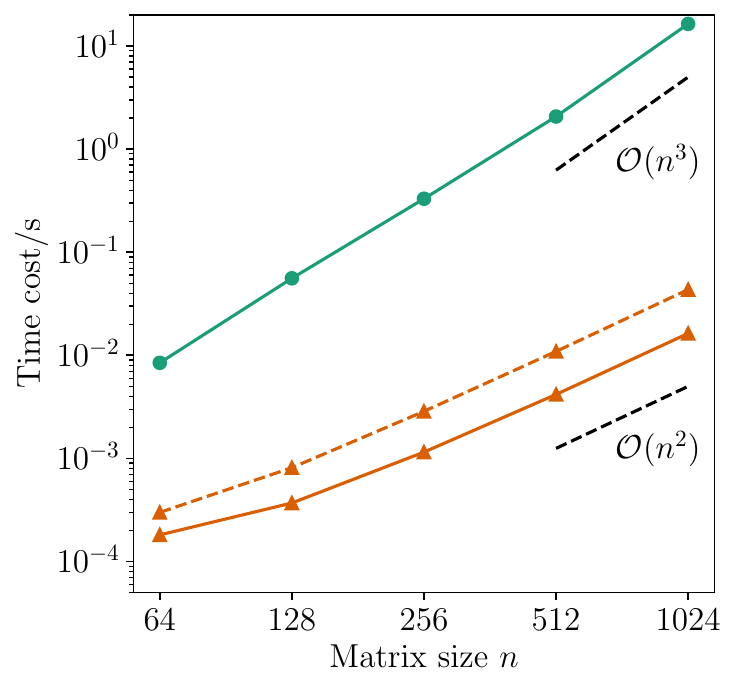}
        \caption{Pfaffian}
    \end{subfigure}
    
    \caption{The scaling of time cost. The direct computation, LRU with matrix inverse updated, and LRU without matrix inverse updated are shown for comparison. Dashed lines indicate reference $\mathcal{O}(n^2)$ and $\mathcal{O}(n^3)$ scalings.}
    \label{fig:one_step}
\end{figure}

We first show the acceleration achieved by LRU on an A100-80GB GPU. To fully exploit the floating-point operations in the GPU, we measure the time cost of 1024 determinants or Pfaffians in parallel implemented by \texttt{jax.vmap}, which mimics batched wavefunction evaluations in modern VMC. In the test of both determinants and Pfaffians, we choose $k=1$ and vary $n$ to show the matrix-size scaling. 

In Fig.\,\ref{fig:one_step}, we plot the time cost of different methods against the matrix size $n$. Although the scaling is unclear for small $n$, for large $n$ we can clearly see that the direct computation of determinants and Pfaffians shows $\mathcal{O}(n^3)$ scaling, while the LRU shows only $\mathcal{O}(n^2)$. The LRU with matrix inverse $\mathbf{A}^{-1}$ updated shows a larger time cost than LRU without matrix inverse updated, but the scaling is the same. Comparing the time cost of the largest presented matrix size $n=1024$, the LRU technique achieves $\sim 200 \times$ acceleration in the determinant and $\sim 1000 \times$ acceleration in the Pfaffian. These results showcase the importance of LRU in the update of determinants and Pfaffians and the great efficiency of \texttt{lrux}.

\subsection{Delayed updates}

\begin{figure}[t]
    \centering
    \includegraphics[width=0.5\linewidth]{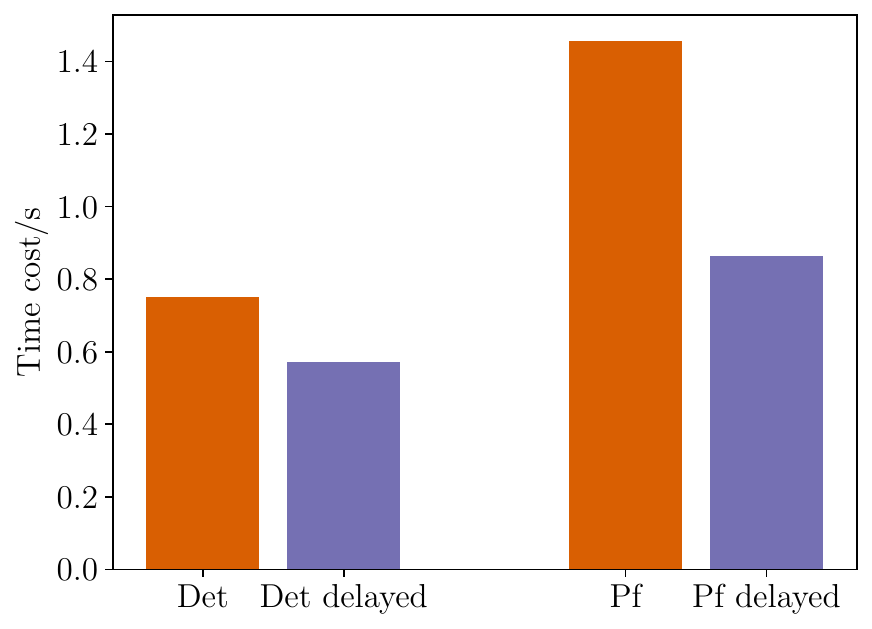}
    \caption{The time cost of LRU with and without delayed updates}
    \label{fig:delayed_updates}
\end{figure}

Next, we compare the efficiency between LRU with matrix inverse directly updated and LRU with delayed updates. The test is still performed on an A100-80GB GPU. We choose a parallel computation of 16384 determinants and Pfaffians with the matrix size $n = 128$, and perform $n$ consecutive updates to compare the total time cost. The upper bound of delayed update steps $T$ is set to maximize the efficiency by performing numerical experiments. After several rounds of tests, we found the optimal choice is $T = 16$ for determinants and $T = 4$ for Pfaffians in our environment.

In Fig.\,\ref{fig:delayed_updates}, we compare the time cost of LRU with and without delayed updates. The efficiency of both methods is in line with each other, while the delayed updates provide an additional $20\% \sim 40\%$ speedup. This improvement could be useful when users want to push the computation to maximum efficiency, especially when LRU is the time bottleneck of the full program. However, the performance of delayed updates relies strongly on the suitable choice of $T$ on users' specific platforms. Therefore, we recommend that users perform their own benchmarks when utilizing the delayed updates. Otherwise, the LRU with direct matrix inverse updates is more appropriate.

\section{Conclusion}

We have introduced \texttt{lrux}, a JAX-based library that provides efficient (delayed) low-rank update algorithms for determinants and Pfaffians, enabling $\mathcal{O}(n^2k)$ scaling for successive updates that commonly arise in QMC and fermionic NQS calculations. By leveraging JAX’s JIT compilation, vectorization, and accelerator support, \texttt{lrux} can achieve substantial speedups on modern GPUs while maintaining numerical stability. The package offers flexible interfaces for dense and one-hot update representations, as well as support for both real and complex data types. We expect lrux to serve as a robust building block for large-scale simulations of interacting fermionic systems and to facilitate the development of next-generation variational wavefunctions and sampling algorithms.

\section*{Acknowledgements}
AC gratefully thanks Garnet Kin-Lic Chan and Markus Heyl for their support. We acknowledge helpful discussions with Anirvan Sengupta, Antoine Georges, Miguel Morales, and Zhou-Quan Wan. We also acknowledge the support and computational resources provided by the Flatiron Institute. The Flatiron Institute is a division of the Simons Foundation.

\begin{appendix}
\numberwithin{equation}{section}

\section{Pfaffian} \label{sec:Pfaffian}

\subsection{Definition}

Pfaffian maps a $2n \times 2n$ skew-symmetric matrix $\mathbf{A}$ to a number $\pf \,\mathbf{A}$. The Pfaffian can be formally defined as follows. Partition all numbers $\{1, ..., 2n \}$ into $n$ pairs $\alpha = \{(i_1, j_1), ..., (i_n, j_n)\}$ with $i_k < j_k$ and $i_1<i_2<...<i_n$, in total $(2n-1)!!$ possible partitions. Then the Pfaffian of matrix $\mathbf{A}$ with elements $A_{i,j}$ is given by
\begin{equation} \label{eq:pf_def}
    \pf \mathbf{A} = \sum_{\alpha} \mathrm{sign}(\alpha) \prod_{k=1}^n A_{i_k, j_k},
\end{equation}
where $\mathrm{sign}(\alpha)$ is the parity of the permutation $(i_1, j_1, i_2, j_2,..., i_n, j_n)$. For example,
\begin{equation}
    \pf \begin{pmatrix}
        0 & a \\ -a & 0
    \end{pmatrix} = a,
\end{equation}
\begin{equation}
    \pf \begin{pmatrix}
        0 & a & b & c \\
        -a & 0 & d & e \\
        -b & -d & 0 & f \\
        -c & -e & -f & 0
    \end{pmatrix} = af - be + cd.
\end{equation}
The time complexity of Pfaffian is $\mathcal{O}(n^3)$, the same as determinant.

In \texttt{lrux}, we define two functions \texttt{pf} and \texttt{slogpf} to compute the Pfaffian of a skew-symmetric matrix. These functions are designed for good numerical stability instead of optimal efficiency under the \texttt{JAX} framework. \texttt{pf(A)} returns the (batched) Pfaffian value directly, while \texttt{slogpf} returns a \texttt{NamedTuple} object containing two values equal to \texttt{jnp.sign(pf(A))} and \texttt{jnp.log(jnp.abs(pf(A)))}.

\subsection{Properties}

For reference, here we list several important properties of the Pfaffian without proof.
\begin{equation} \label{eq:pf2=det}
        \pf^2(\mathbf{A}) = \det(\mathbf{A}),
    \end{equation}
\begin{equation} \label{eq:pf->det_pf}
    \pf (\mathbf{BAB}^T) = \det(\mathbf{B}) \pf(\mathbf{A}),
\end{equation}
\begin{equation} \label{eq:pf_offdiagonal}
    \pf \begin{pmatrix}
        \mathbf{0} & \mathbf{A} \\
        \mathbf{-A}^T & \mathbf{0}
    \end{pmatrix}
    = (-1)^{n(n-1)/2} \det(\mathbf{A}),
\end{equation}
\begin{equation}  \label{eq:pf_diagonal}
    \pf \begin{pmatrix}
        \mathbf{A} & \mathbf{0} \\ \mathbf{0} & \mathbf{A}'
    \end{pmatrix}
    = \pf(\mathbf{A}) \pf(\mathbf{A}'),
\end{equation}
\begin{equation} \label{eq:pf_low_rank}
    \frac{\pf(\mathbf{A} + \mathbf{BCB}^T)}{\pf(\mathbf{A})}
    = \frac{\pf(\mathbf{C}^{-1} + \mathbf{B}^T \mathbf{A}^{-1} \mathbf{B})}{\pf(\mathbf{C}^{-1})},
\end{equation}
\begin{equation} \label{eq:pf_block}
    \pf\begin{pmatrix}
        \mathbf{M} & \mathbf{Q} \\
        -\mathbf{Q}^T & \mathbf{N}
    \end{pmatrix} 
    = \pf(\mathbf{M}) \, \pf(\mathbf{N} + \mathbf{Q}^T \mathbf{M}^{-1} \mathbf{Q})
    = \pf(\mathbf{N}) \, \pf(\mathbf{M} + \mathbf{Q} \mathbf{N}^{-1} \mathbf{Q}^T).
\end{equation}

\subsection{Gradient}

Here, we derive the gradient of the Pfaffian. Assume a variation $\mathrm{d}x$ is imposed on the $i$'th row and $j$'th column of $\mathbf{A}$. Then
\begin{equation}
    (\mathrm{d}\mathbf{A})_{kl} = \delta_{ik} \delta_{jl} \, \mathrm{d}x.
\end{equation}
Due to the skew-symmetry constraint, we skew-symmetrize the matrix $\mathbf{A}$ to obtain
\begin{equation}
    \mathbf{A}' = \frac{\mathbf{A} - \mathbf{A}^T}{2},
\end{equation}
whose variation is
\begin{equation}
    (\mathrm{d}\mathbf{A}')_{kl} = \frac{\delta_{ik} \delta_{jl} - \delta_{il} \delta_{jk}}{2} \mathrm{d}x.
\end{equation}
Then we have $\mathbf{A'}^T = -\mathbf{A'}$ and $\mathrm{d}\mathbf{A'}^T = -\mathrm{d}\mathbf{A}'$. The value of $\pf \mathbf{A}'$ after a variation is given by
\begin{equation}
\begin{split}
    \pf (\mathbf{A}' + \mathrm{d}\mathbf{A}')
    &= \pf \left[ \left(\mathbf{1} + \frac{1}{2}\mathbf{A'}^{-1} \mathrm{d}\mathbf{A'} \right)^T \mathbf{A'} \left( \mathbf{1} + \frac{1}{2} \mathbf{A'}^{-1} \mathrm{d}\mathbf{A}' \right) \right] + \mathcal{O}(\mathrm{d}x^2) \\
    &= \pf(\mathbf{A}') \det \left( \mathbf{1} + \frac{1}{2} \mathbf{A'}^{-1} \mathrm{d}\mathbf{A}' \right) + \mathcal{O}(\mathrm{d}x^2) \\
    &= \pf(\mathbf{A}') \left(1 + \frac{1}{2}\mathrm{tr}(\mathbf{A'}^{-1} \mathrm{d}\mathbf{A}') \right) + \mathcal{O}(\mathrm{d}x^2), \\
\end{split}
\end{equation}
where we have utilized the skew-symmetry property of $\mathbf{A}'$ and Eq.\,\eqref{eq:pf->det_pf}. With the Einstein summation rule, the trace above can be computed as
\begin{equation}
    \mathrm{tr}(\mathbf{A'}^{-1} \mathrm{d}\mathbf{A}')
    = (\mathbf{A'}^{-1})_{lk} (\mathrm{d}\mathbf{A}')_{kl}
    = (\mathbf{A'}^{-1})_{lk} \frac{\delta_{ik} \delta_{jl} - \delta_{il} \delta_{jk}}{2} \mathrm{d}x
    = (\mathbf{A'}^{-1})_{ji} \, \mathrm{d}x.
\end{equation}
Then the partial derivative of $\pf \mathbf{A}'$ can be computed as
\begin{equation} \label{eq:pf_grad}
    \frac{\partial \, \pf \mathbf{A'}}{\partial A_{ij}}
    = \lim_{\mathrm{d}x \to 0} \frac{\pf(\mathbf{A'} + \mathrm{d}\mathbf{A'}) - \pf\mathbf{A'}}{\mathrm{d}x}
    = \frac{1}{2} \pf (\mathbf{A}') (\mathbf{A'^{-1}})_{ji}.
\end{equation}
In \texttt{lrux}, the gradients of functions \texttt{pf} and \texttt{slogpf} are customized based on Eq.\,\eqref{eq:pf_grad} using \texttt{jax.custom\_jvp}, and the real and complex data types are both supported.

\end{appendix}

\bibliography{reference}

\end{document}